\documentclass[prl,aps,showpacs,twocolumn,superscriptaddress,floatfix]{revtex4-1}

\usepackage{epsfig}
\usepackage{amsmath}
\usepackage{amssymb}
\usepackage{amsfonts}
\usepackage{mathptmx}
\usepackage{eucal}
\usepackage{bm}
\usepackage{color}
\usepackage[colorlinks,linkcolor=blue,citecolor=blue]{hyperref}

\usepackage{epstopdf}

\usepackage{graphicx}

\usepackage{natbib}

\begin{document}
\title{Phase-tunable colossal  magneto-heat resistance in ferromagnetic Josephson thermal valves}
\author{F. Giazotto}
\email{giazotto@sns.it}
\affiliation{NEST, Instituto Nanoscienze-CNR and Scuola Normale Superiore, I-56127 Pisa, Italy}
\author{F. S. Bergeret}
\email{sebastian\_bergeret@ehu.es}
\affiliation{Centro de F\'{i}sica de Materiales (CFM-MPC), Centro
Mixto CSIC-UPV/EHU, Manuel de Lardizabal 4, E-20018 San
Sebasti\'{a}n, Spain}
\affiliation{Donostia International Physics Center (DIPC), Manuel
de Lardizabal 5, E-20018 San Sebasti\'{a}n, Spain}
\affiliation{Institut f\"ur Physik, Carl von Ossietzky Universit\"at, D-26111 Oldenburg, Germany}

\begin{abstract}
We propose a heat valve based on  the interplay between \emph{thermal} transport and proximity-induced \emph{exchange} splitting in Josephson tunnel junctions.
We demonstrate that the junction heat conductance strongly depends on the relative alignment of the exchange  fields induced in the superconductors.
\emph{Colossal} magneto-heat resistance ratios as large as $\sim10^7\%$ are predicted to occur under proper temperature and phase conditions, as well as suitable ferromagnet-superconductor combinations.  Moreover, the quantum phase tailoring, intrinsic to  the Josephson coupling,  offers an  additional degree of freedom for the  control of the heat conductance. Our predictions for
 the phase-coherent and spin-dependent  tuning of the  thermal flux can provide a useful tool  for heat management at the nanoscale.  
\end{abstract}

\maketitle

The study of heat transport and dynamics in meso- \cite{Giazotto2006} and nanoscopic \cite{Dubi2011} solid-state systems, is a research field that has attracted much attention in recent years because of  the impressive progress achieved  in nanoscience and nanofabrication techniques.  At such scale heat  may play a significant role in determining the  properties of  the devices,  and therefore it is of particular interest to control and manipulate \cite{heattransistor,ser}  the thermal flux as well to 
understand the  origin of   dissipative phenomena.  
Prototypical cases in which the understanding of heat transport is crucial are, for instance, the fine temperature control in ultrasensitive cryogenic radiation detectors \cite{Giazotto2006},  general cooling applications at the nanoscale \cite{Giazotto2006}, and  the emerging field of   \emph{coherent} caloritronic circuitry where the quantum phase allows for enhanced operation \cite{Meschke2006,Vinokur2003,Eom1998,Chandrasekhar2009,Ryazanov1982,Panaitov1984,virtanen2007}. 

It has been known for a few decades that phase-dependent thermal  transport  through  weakly-coupled superconducting condensates is in principle possible \cite{Maki1965,Guttman97,Guttman98,Zhao2003,Zhao2004}.   However, only recently the first Josephson heat interferometer was demonstrated \cite{giazotto2012,martinez2012,giazottoexp2012}.
The experiment of Ref. \cite{giazottoexp2012}  proves that, in addition  to  the Josephson  charge supercurrent, phase coherence extends  to dissipative observables such as the thermal current.  This heat interferometer  represents a  prototypical building block to implement  future coherent caloritronic circuits like, for instance, heat transistors and thermal splitters.   

In this Letter we put forward the concept of a ferromagnetic Josephson junction acting as a thermal valve. 
In particular, we address the interplay between thermal transport and proximity-induced \emph{exchange} splitting in a Josephson tunnel weak-link consisting of two superconducting electrodes  with an internal exchange splitting. 
The latter is induced   from nearby-contacted ferromagnetic layers [see Fig. 1(a)].
 We show that the junction thermal conductance strongly depends on the relative alignment of the exchange fields induced in the superconductors. As a results, \emph{colossal} magneto-heat resistance ratios  as large as $\sim10^7\%$ are predicted to occur for suitable exchange fields and proper temperature conditions. Moreover,  the quantum phase tailoring, characteristic  for  the Josephson effect, 
 adds a further degree of freedom for enhanced heat conductance control. 

\begin{figure}[tb]
\includegraphics[width=\columnwidth]{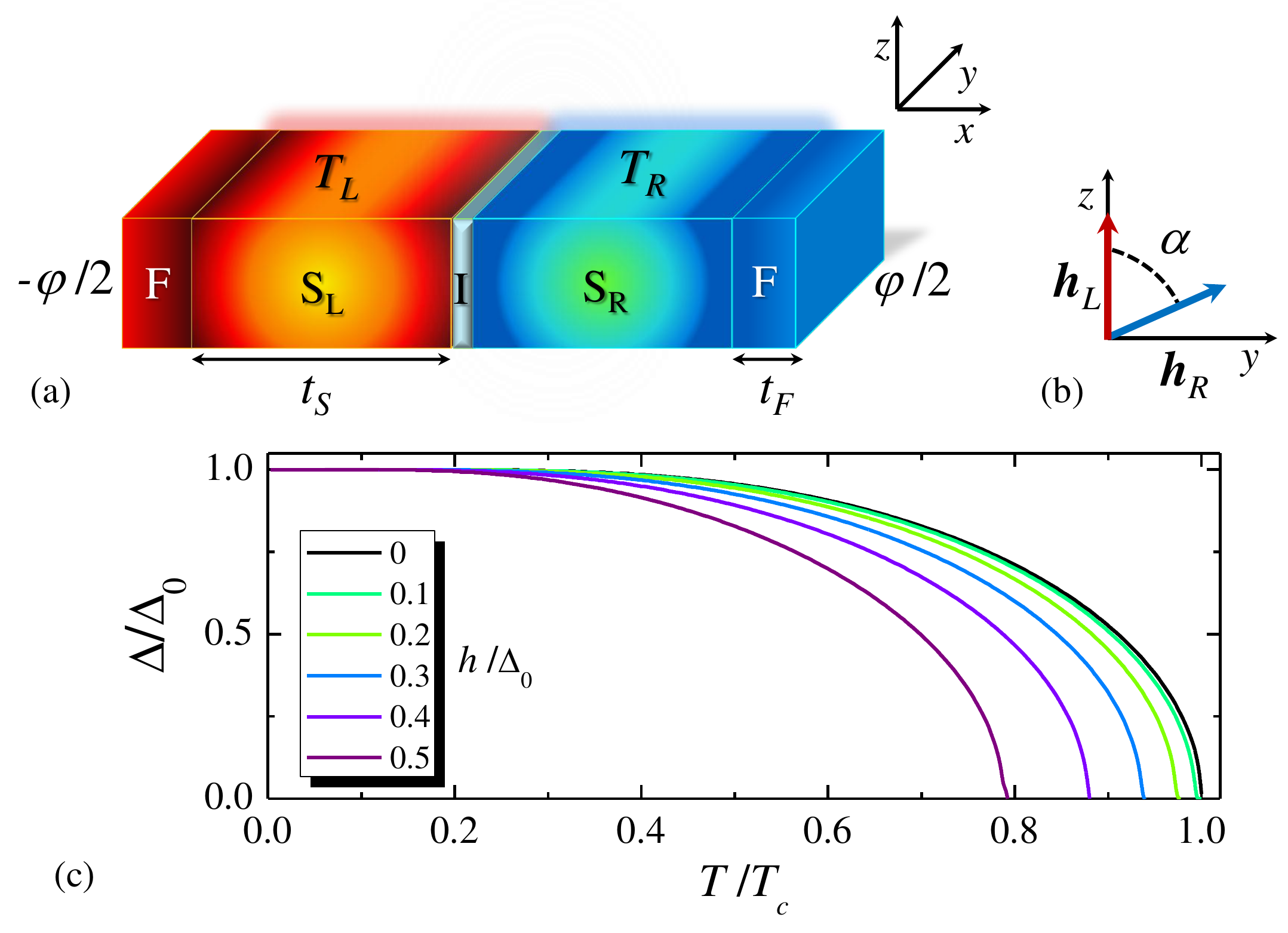}\vspace{-2mm}
\caption{(Color online) (a) A schematical view  of the FSISF  Josephson heat valve discussed in the text. 
(b) The exchange fields ($\emph{\textbf{h}}_{L,R}$) in the F layers are confined to the $z-y$ plane, and are misaligned by an angle $\alpha$.
(c) Temperature dependence of the self-consistently calculated  superconducting order parameter $\Delta$  for different values of the exchange field $h$. 
$\Delta_0$ is the zero-temperature, zero-exchange field order parameter and $T_c$ is the superconducting critical temperature.}
\label{fig1}
\end{figure}

Our system is schematized in Fig. 1(a). It consists of two equal ferromagnet-superconductor bilayers (FS$_{\text{L,R}}$) tunnel-coupled through an insulating barrier (I) and implementing a Josephson junction. The FS$_\text{L}$ and FS$_\text{R}$ bilayers are in thermal steady-state and reside at different temperatures $T_L$ and $T_R$, respectively.
For definiteness, we assume $T_L\geq T_R$ so that the structure is temperature-biased only, while there is no voltage drop across the Josephson junction. 
$t_S$ ($t_F$) labels the S  (F) layer thickness while $\varphi$ denotes the macroscopic quantum phase difference over the junction. Furthermore, the $z$-axis is the one parallel to the magnetization (exchange field) of the left F layer ($\emph{\textbf{h}}_L$), which is kept fixed, whereas the one in the right ferromagnet ($\emph{\textbf{h}}_R$) is misaligned by an angle $\alpha$ [see Fig. 1(b)]. Experimentally this can be achieved either by using ferromagnetic films with different coercive fields or by pinning the magnetization in the left electrode through an exchange-bias with an additional magnetic layer \cite{exchangebias}. 
 $\emph{\textbf{h}}_R$ can therefore be freely rotated by applying an in-plane magnetic field as low as a few tens of Oe.

We first derive an expression for the heat current ($\dot Q$) flowing through the Josephson junction. The latter  can be expressed  in terms of  the quasiclassical Green's functions  (GFs)  of the left and right electrodes
\begin{equation}
\dot Q=\frac{1}{16e^2R_N}\int  \epsilon {\rm Tr}\left\{\left[G_R,G_L\right]^K\right\}d\epsilon\; .
\label{dQ}
\end{equation}
Here the trace  is taken over the spin$\otimes$particle-hole space
while the  functions  $\check G_{R(L}) $  are $8\times8$ matrices in the Keldysh$\otimes$particle-hole$\otimes$spin space:
\begin{equation}
G_j=\left(
\begin{array}{cc}
\check g_j^R & \check g_j^K\\
0 & \check g_j^A\\
\end{array}
\right)\; ,
\end{equation}
where $j=R,L$.  
The symbols $\check .$ and $\hat .$ denote 4$\times$4 matrices in particle-hole$\otimes$spin and $2\times2$ matrices in spin-space, respectively. Furthermore, $R_N$ is the normal-state resistance of the junction and $e$ is the electron charge.

We assume that the electrodes are in thermal equilibrium, thus  the Keldysh component of the GFs is given by $\check g_j^K(\epsilon)=(\check g_j^R-\check g_j^A)F_j$, where $F_j=\tanh[\epsilon/(2T_j)]$ is the electronic distribution function,  and  $T_j$ is the temperature of  the $j$ electrode. According to Eq. (\ref{dQ}) there is a finite  heat current  flowing through the junction if $T_R\neq T_L$ which is given by 
 \begin{equation}
\dot Q=\frac{1}{2e^2R_N}\int d\epsilon. \epsilon {\rm Tr}\left[\hat N_L\hat N_R-\hat M_L\hat M_R\cos\varphi\right]\left[ F_R-F_L\right]\;.
\label{dQ2}
\end{equation}
The two contributions to the heat current stem from the normal,   $\hat N_j=(\hat g_j^R-\hat g_j^A) /2$, and  phase-coherent (anomalous), $\hat M_j=(\hat f_j^R-\hat f_j^A) /2$, parts of the quasiparticle  spectral function \cite{Maki1965,Zhao2003}.  Equation (\ref{dQ2}) is the generalization of the Maki-Griffin heat current equation \cite{Maki1965} for the case of spin-dependent density of states (DoS).
In particular, we obtain  the  oscillatory behavior of the heat current as a function of the superconducting phase difference $\varphi$  predicted for the first time in Ref. \cite{Maki1965}, and recently demonstrated in Ref. \cite{giazottoexp2012}.  
We stress that a pure temperature bias across the junction is a crucial condition to preserve phase dependence in thermal transport. Indeed, 
any voltage drop occurring across the Josephson weak-link would make $\varphi$ time-dependent and, therefore, the $\varphi$-dependent component of $\dot Q$ in Eq. (\ref{dQ2}) would not contribute to the DC heat transport \cite{Maki1965,Guttman98,giazottoexp2012}. 

Instead of analyzing the heat current, that depends on a generic temperature difference across the junction, we shall focus on the behavior of the  
thermal conductance ($\kappa$) which is defined for small temperature differences as
\begin{equation}
\kappa=\frac{\dot Q}{\delta T}=-\frac{1}{2e^2R_N}\int d\epsilon. \epsilon  \left(\frac{\partial F}{\partial T}\right)  {\rm Tr}\left[\hat N_L\hat N_R-\hat M_L\hat M_R\cos\varphi\right]\; ,
\label{G}
\end{equation}
where $\delta T=T_L-T_R$, and $(\partial F/\partial T)=-\epsilon/[2T^2\cosh^2(\epsilon/2T)]$.  By deriving the second equality we have assumed that $\delta T\ll T=(T_R+T_L)/2$.
Equations (\ref{dQ}) and (\ref{G}) are  rather general, and allow to compute the heat current  and the thermal conductance for an arbitrary tunneling junction provided that values of the GFs on both side of the interfaces are known.
\begin{figure}[tb]
\includegraphics[width=\columnwidth]{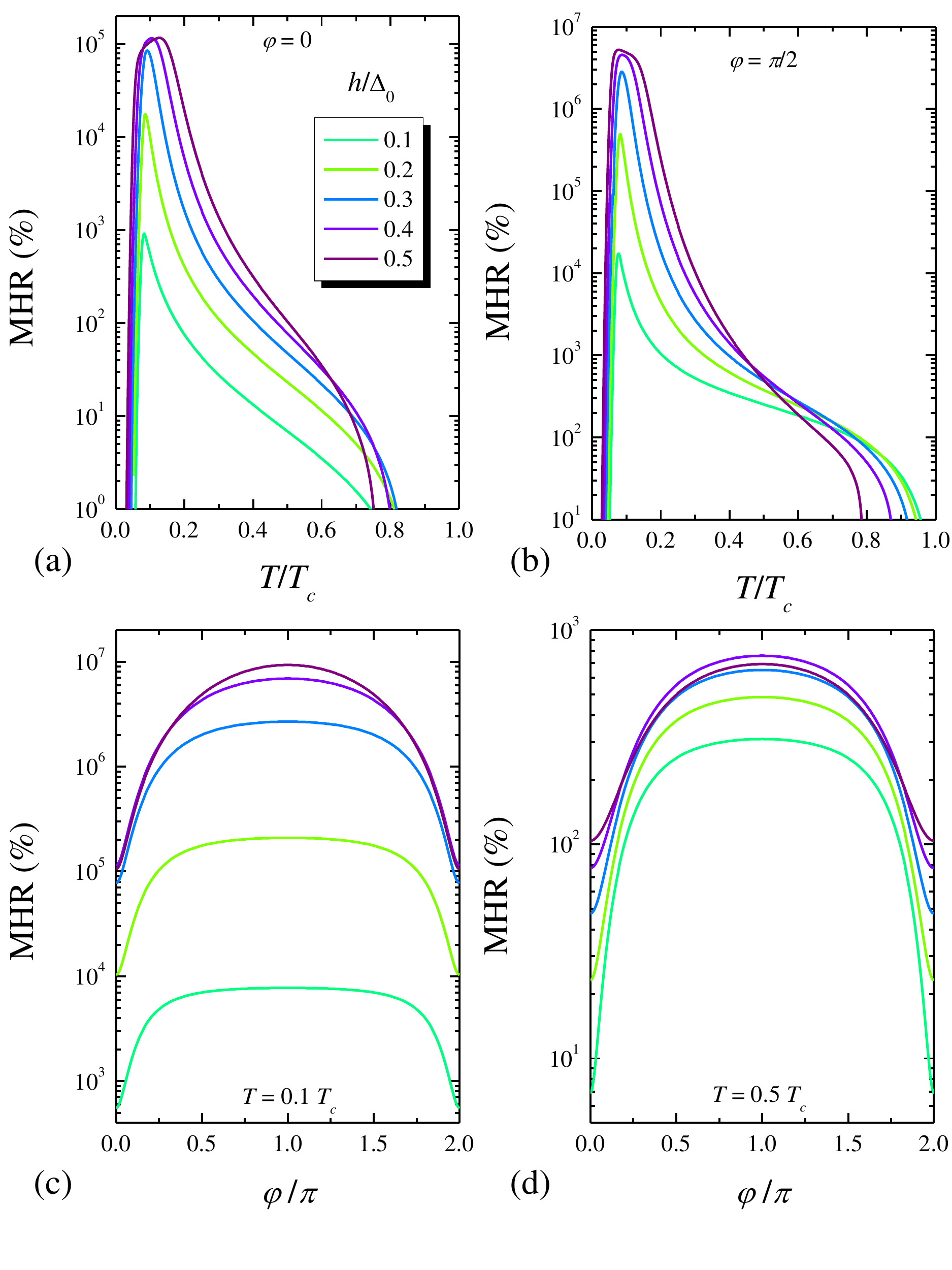} \vspace{-2mm}
\caption{(Color online) (a)  Magneto-heat resistance ratio MHR vs temperature $T$ calculated for a few values of the exchange field $h$ at $\varphi=0$.
(b) MHR ratio vs $T$ calculated for the same values of $h$ as in panel (a) at $\varphi=\pi/2$.
(c) MHR ratio vs $\varphi$ calculated for several values of the exchange field at $T=0.1T_c$.
(d) MHR ratio vs $\varphi$ calculated at $T=0.5T_c$  for the same values of $h$ as in panel (c).
}
\label{fig2}
\end{figure}

With the help of Eq. (\ref{G}) we can determine  the heat conductance for the junction sketched  in Fig. 1(a).
 We assume that $|\emph{\textbf{h}}_L|=|\emph{\textbf{h}}_R|=h$, and that the S/F interface is highly transmissive so that both the superconductor and the ferromagnet  are  strongly affected by the \emph{proximity effect} \cite{Buzdin2005,Sebas2005}.  
At the same time, in order to preserve superconductivity in the leads,  we assume that the F layers are thin enough.  
In particular, if the thickness $t_{S(F)}$ of the  superconducting (ferromagnetic) layer is  smaller than the characteristic  length over which the GFs  vary, one can   integrate the quasiclassical equations over the thickness  of the S/F bilayers \cite{Bergeret2001}. 
After such procedure one obtains  for the retarded and advanced GFs $\check g_{R(L)}^{R(A)}=\hat g_{R(L)}^{R(A)}\tau_3+\hat f_{R(L)}^{R(A)}(i\tau_1\cos\varphi/2\pm i\tau_2\sin\varphi/2)$, 
where $\tau$'s are the Pauli matrices in particle-hole space.  We focus first on the case that  the  magnetizations of the F layers in Fig. 1 are either parallel (P) or antiparallel (AP) to each other. 
Thus GFs   $\hat g^{R(A)}$ and $\hat f^{R(A)}$ are $2\times2$ diagonal matrices  in spin space with  diagonal  elements   given by \cite{Bergeret2001}
\begin{eqnarray}
g_\pm^{R}=\frac{\epsilon\pm h}{\sqrt{(\epsilon\pm h+i\Gamma)^2-\Delta^2(h,T)}}\label{GF1}\\
f_\pm^{R}= \frac{\Delta(T)}{\sqrt{(\epsilon\pm h+i\Gamma)^2-\Delta^2(h,T)}},
\label{GFs}
\end{eqnarray}
where $h$ and $\Delta$ are the effective values of the exchange field and superconducting order parameter  in the S/F bilayer, respectively. 
In particular, $\Delta$ has to be determined self-consistently. The temperature dependence of the order parameter for different values of $h$ is shown in Fig. 1(c).
The parameter $\Gamma$  in Eqs. (5-6) accounts for  the inelastic scattering energy rate within the relaxation time approximation \cite{Dynes1984,Pekola2004,Gamma}.  
Similar expressions hold for the advanced GFs   by replacing in Eqs. (5-6) $i\Gamma$ by $-i\Gamma$.  The real part of the functions $g^R_\pm$ gives the modified DoS in the superconductors which is spin-dependent due to the finite exchange field in the F layers.

 The heat  conductance is thus obtained from Eq. (\ref{G}) 
\begin{equation}
\kappa_X=-\frac{1}{2e^2R_N}\int  d\epsilon \epsilon \left(\frac{\partial F}{\partial T}\right) {\mathcal A}_X(\epsilon),\label{GP}
\end{equation}
where $X=P, AP$, ${\mathcal A}_P(\epsilon)=\sum_{\alpha=\pm}\left[\hat N_{L\alpha}\hat N_{R\alpha}-\hat M_{L\alpha}\hat M_{R\alpha}\cos\varphi\right]$ and 
${\cal A}_{AP}(\epsilon)=2\left[\hat N_{L+}\hat N_{R-}-\hat M_{L+}\hat M_{R-}\cos\varphi\right]$.
We propose an experiment in which  one can switch between the P and AP configurations, and  determine the  magneto-heat resistance (MHR) ratio defined as
\begin{equation}
{\rm MHR}=\frac{\kappa_P-\kappa_{AP}}{\kappa_{AP}}.
\label{MHR}
\end{equation}

 In Fig. 2 we show the  behavior of the MHR as a function  of temperature and the superconducting phase difference.
 All panels show an overall  huge MHR ratio ($\sim 10^5-10^7\%$) within a broad range of parameters.   
 We  demonstrate in this way that by  switching  between the P and AP configuration one realizes an almost \emph{perfect} heat valve effect   as the thermal conductance in the AP configuration is practically negligible with respect to that in the P one. 
 This colossal MHR is  one of the key results of the present letter.
Figures 2(a) and 2(b) show that  the heat valve  effect is maximized at certain  finite  temperature (i.e., for $T/T_c\sim 0.1$) and  for sufficiently large exchange fields. Here $T_c$ is the superconducting critical temperature. 
It is worth emphasizing that  due to the $\cos\varphi$ interference term in Eq. (\ref{G}) the MHR ratio can be additionally largely tuned by the  phase difference between the superconductors. 
Such a phase-tunable thermal transport mechanism originates from the Josephson effect and is unique to weakly-coupled superconductors \cite{Maki1965}. 
In the lower panels of Fig. 2  the MHR dependence on $\varphi$ is displayed. 
The  minimum value of the MHR is achieved for zero phase difference, whereas it reaches its maximum value for $\varphi=\pi$. 
We also emphasize that the phase-coherent term in Eq. (\ref{G}) does not describe pure tunneling of Cooper pairs \cite{Maki1965,Guttman97}. 
Furthermore, we point out that while the P configuration maximizes the heat current, the DC Josephson effect is maximized by the AP one \cite{Bergeret2001}.

\begin{figure}[tb]
\includegraphics[width=\columnwidth]{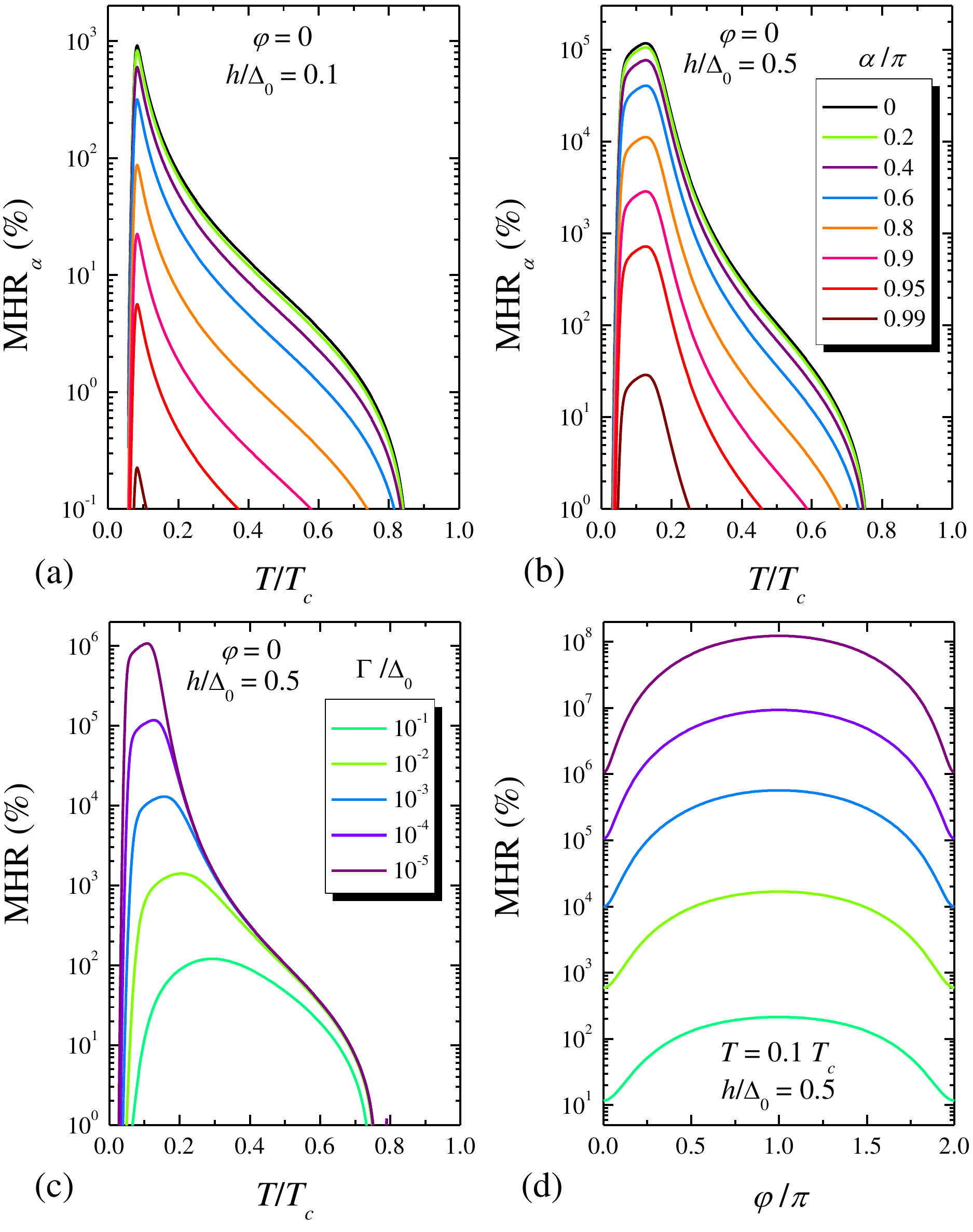} \vspace{-2mm}
\caption{(Color online) (a) MHR$_{\alpha}$ ratio vs $T$ calculated for several values of the misalignment angle $\alpha$ at $h=0.1 \Delta_0$ and $\varphi=0$. 
(b) MHR$_{\alpha}$ ratio vs $T$ calculated for the same $\alpha$ values as in panel (a) at  $h=0.5 \Delta_0$ and $\varphi=0$.
(c) MHR ratio vs $T$ calculated for a few values of $\Gamma$ at $h=0.5 \Delta_0$ and $\varphi=0$.
(d) MHR ratio vs $\varphi$ calculated fot the same $\Gamma$ values as in panel (c) at $h=0.5 \Delta_0$ and $T=0.1 T_c$. }
\label{fig3}
\end{figure}
The  obtained colossal  MHR  ratio can be understood by inspection of   Eq. (\ref{GP}).  If we assume for simplicity that   $\varphi=\pi/2$ [see Fig. 1(b)] then   only the normal GFs contribute to ${\mathcal A}_X$ [{\it cf.} Eq. (\ref{G})].  The heat current (and hence the thermal conductance) is a consequence of quasiparticle transmission from the hot to the cold electrode.  For a given energy, the number of states available for the heat transport is given by the   
 spectral function ${\mathcal A}_X$ which is the product of the DoS  on both side of the tunnel barrier.
Due to the exchange field, the DoS is spin-dependent, and shows a BCS-like shape with spin-dependent energy gap at $\Delta_\pm=\Delta\pm h$ [see Eq. (\ref{GF1})], equivalent to a Zeeman-split superconductor in a magnetic field \cite{Meservey1994}. 
 In the P configuration,  the  DoS of the left and right electrode  coincide for both spin-up and spin-down,  and therefore quasiparticles with  energies around  $\epsilon\sim\Delta_\pm$  contribute at most  to the heat conductance \cite{foot2}.
 The situation is different in  the AP configuration, where the DoS for each spin-channel is shifted on both side of the barrier by an amount $2h$.  
The main contribution to $\kappa_{AP}$ comes from quasiparticles with energies $\epsilon\sim\Delta_+$ and ${\cal A}_{AP}$  is approximately  a factor $\sim\sqrt{\Gamma/\Delta}$ smaller than ${\cal A}_{P}$.
  Moreover, in both the P and AP cases the contribution from  ${\mathcal A}_{P(AP)}$ is weighted by the function $\epsilon\partial F/\partial T(\epsilon)$. The latter decays as $e^{-\epsilon/2T}$ for $\epsilon>2T$ and hence the main contribution to $\kappa_{AP}$ in Eq. (\ref{GP}) (from  $\epsilon\sim\Delta_+$) has an additional exponentially small  factor $e^{-h/T}$ with respect to the main contribution to  $\kappa_P$ (from  $\epsilon\sim\Delta_-$).
All of this explains the smallness of $\kappa_{AP}$ and the huge  MHR ratio obtained for  sufficiently large values of the  exchange field.  
   
As discussed above, the maximum  MHR ratio is  reached for a certain  finite temperature. According to Figs. 2(a,b) a further increase of $T$ leads to a decrease of the MHR, which can be explained, on the one hand,  by  the suppression of the energy gap $\Delta(T)$ and on the other hand,  by the fact that increasing $T$ the contribution from quasiparticles with energies larger than $\Delta_\pm$ becomes  more and more important leading to a smaller  difference between $\kappa_{AP}$ and $\kappa_P$. 
 Notice that for $0\leq\varphi< \pi/2$ the condensate part of the spectral function ${\mathcal A}_X(\epsilon)$ [see Eq. (\ref{GP})] gives a negative contribution to the heat conductance.
This explains the lower values of MHR for small phase difference shown in Figs. 2(c) and 2(d).

For  an arbitrary angle $\alpha$  between the magnetizations of the left and right electrode [see Fig. 1(b)] we define   MHR$_\alpha$  as
 ${\rm MHR_\alpha}=(\kappa_\alpha-\kappa_{AP})/{\kappa_{AP}}$, where
$\kappa_\alpha=\kappa_P\cos^2(\alpha/2)+\kappa_{AP}\sin^2(\alpha/2)$.
The last expression  for  $\kappa_\alpha$ can be obtained
 straightforwardly from Eq. (\ref{dQ}) by rotating the right Green function according to $\check G_R=\check R_\alpha \check G_0\check R_\alpha^\dagger$, where $\check G_0$ is the GF for the case $\alpha=0$ and $R_\alpha=\exp[i\tau_3\sigma_1\alpha/2]$ \cite{Bergeret2001b,Bergeret2012}. 
 In Figs. 3(a) and 3(b) we show the temperature dependence of MHR$_\alpha$ for different values of $\alpha$ at $\varphi=0$.   
All curves show similar behavior, and again very large values for the MHR can be achieved with a proper choice of  the parameters.   According to Figs. 3(a) and 3(b), the effect is maximized for  $\alpha=0$, i.e., when the junction is switched between the P and AP configurations.  
Figures 3(c) and 3(d) show  the impact of the inelastic parameter $\Gamma$ on the MHR.  The overall tendency is that by increasing $\Gamma$ the MHR ratio is reduced, as the ``normal'' character of transport is strengthened in the heat valve leading to a suppression of the large MHR ratio. The latter, indeed, originates from the presence of the superconducting gap. 
Moreover, as displayed in Fig. 3(c), the MHR ratio reaches its maximum at higher temperature by increasing $\Gamma$. 

In  light of a realistic  implementation of the present heat valve, soft ferromagnetic alloys such as Cu$_{1-x}$Ni$_x$ \cite{Ryzanov2001} or Pd$_{1-x}$Ni$_x$ \cite{Kontos2001}, which allow fine tuning of the exchange field through a suitable choice of $x$, combined with a conventional superconductor ({\it e.g.} aluminum or niobium)  might be suitable candidates.
We note that all the results presented above have been obtained assuming  highly-transparent S/F interfaces. 
But nevertheless, they are qualitatively valid as well in the case of a finite S/F interface resistance $R_b$. 
In such a case  the superconductor still exhibits a  spin-split DoS,  however with an additional damping factor determined by $R_b$. The latter will suppress the MHR similarly as it does a finite $\Gamma$. 
Furthermore, according to our model one can also design the heat valve of Fig. 1(a) by using ferromagnetic insulators (FIs), as for example Eu chalcogenides barriers,  instead of metallic ferromagnets. 
In such a case it was experimentally proved \cite{Moodera90,Moodera08,Miao2009} that the DoS in the superconductor is modified and shows the spin-splitting  needed to obtain the heat valve effect. 
Therefore,  all the conclusions drawn above remain valid if one designs the junction by exploiting  FIs for the F layers.

With regards to  potential applications,  the present thermal valve  can be used  whenever a precise control and mastering of the temperature is required, for instance, for on-chip heat management as a \emph{switchable heat sink}. 
This setup can  be useful as well  to  tune the operation temperature of sensitive radiation detectors \cite{Giazotto2006,Giazotto2008}. 
In the context of quantum computing architectures \cite{Nielsen2002}  the Josephson thermal valve can also be used to influence the behavior and the dynamics of two-level quantum systems through temperature manipulation. 
Similarly, the relation between Josephson critical supercurrent and the temperature can be exploited for designing tunable thermal Josephson weak-links of different kinds \cite{Giazotto2006,Giazotto2004,Tirelli2008}.

In conclusion, we have investigated thermal transport through a heat valve consisting of a Josephson junction between two S/F bilayers as electrodes.
In particular, we predict that the heat conductance depends strongly on the the relative alignment of the magnetizations of the F layers. 
Under specific conditions of temperature bias and phase difference across the junction one can obtain a colossal magneto-heat resistance ratio as high as several orders of magnitude. 
The spin-dependent and phase-tunable mechanisms  of heat flux control discussed in this letter  will likely prove useful for  thermal management at the nanoscale,  and  for the development of coherent spin caloritronic nanocircuits \cite{bauer2010,bauer2012} .

{\it Acknowledgements.-} F.G. acknowledges the FP7 program No. 228464 ``MICROKELVIN'', the Italian Ministry of Defense through the PNRM project ``TERASUPER'', and  the Marie Curie Initial Training Action (ITN) Q-NET 264034
for partial financial support. 
The work of F.S.B  was supported by the Spanish Ministry of Economy
and Competitiveness under Project FIS2011-28851-C02-02. F.S.B thanks Prof. Martin
Holthaus and his group for their kind hospitality at the Physics Institute of the 
Oldenburg University.


\begin{thebibliography}{99}
\bibitem{Giazotto2006} 
	F. Giazotto, T. T. Heikkil\"a, A. Luukanen, A. M. Savin, and J. P. Pekola, Rev. Mod. Phys. \textbf{78}, 217 (2006).
\bibitem{Dubi2011} 
	Y. Dubi and M. Di Ventra, Rev. Mod. Phys. \textbf{83}, 131 (2011). 
\bibitem{heattransistor} 
	O.-P. Saira, M. Meschke, F. Giazotto, A. M. Savin, M. M\"ott\"onen, and J. P. Pekola, Phys. Rev. Lett. \textbf{99}, 027203 (2007).
\bibitem{ser} 
	J. P. Pekola, F. Giazotto, and O.-P. Saira, Phys. Rev. Lett. \textbf{98}, 037201 (2007).
\bibitem{Meschke2006}
	M. Meschke, W. Guichard, and J. P. Pekola, Nature \textbf{444}, 187 (2006).
\bibitem{Vinokur2003}
	E. V. Bezuglyi and V. Vinokur, Phys. Rev. Lett. \textbf{91}, 137002 (2003).
\bibitem{Eom1998}
	J. Eom, C.-J. Chien, and V. Chandrasekhar, Phys. Rev. Lett. \textbf{81}, 437 (1998).
\bibitem{Chandrasekhar2009}
	V. Chandrasekhar, Supercond. Sci. Technol. \textbf{22}, 083001 (2009).
\bibitem{Ryazanov1982}
	V. V. Ryazanov and V. V. Schmidt, Solid State Commun. \textbf{42}, 733 (1982).
\bibitem{Panaitov1984}
	G. I. Panaitov, V. V. Ryazanov, and V. V. Schmidt, Phys. Lett. \textbf{100}, 301 (1984).
\bibitem{virtanen2007}
	P. Virtanen and T. T. Heikkil\"a, Appl. Phys. A \textbf{89}, 625 (2007). 
\bibitem{Maki1965} 
	K. Maki and A. Griffin, Phys. Rev. Lett. {\bf 15}, 921 (1965).
\bibitem{Guttman97} 
	G. D. Guttman, B. Nathanson, E. Ben-Jacob, and D. J. Bergman, Phys. Rev. B \textbf{55}, 3849 (1997).
\bibitem{Guttman98} 
	G. D. Guttman, E. Ben-Jacob, and D. J. Bergman, Phys. Rev. B \textbf{57}, 2717 (1998).
\bibitem{Zhao2003} 
	E. Zhao, T. L\"oftwander, and J. A. Sauls, Phys. Rev. Lett. \textbf{91}, 077003 (2003).
\bibitem{Zhao2004} 
	E. Zhao, T. L\"oftwander, and J. A. Sauls, Phys. Rev. B \textbf{69}, 134503 (2004).
\bibitem{giazotto2012} 
	F. Giazotto and M. J. Mart\'inez-P\'erez, Appl. Phys. Lett. \textbf{101}, 102601 (2012).
\bibitem{martinez2012} 
	M. J. Mart\'inez-P\'erez and F. Giazotto, submitted (2012) arXiv:1210.7187v1.
\bibitem{giazottoexp2012} 
	F. Giazotto and M. J. Mart\'inez-P\'erez, Nature, in print (2012).
\bibitem{exchangebias} 
J. Nogu\'{e}s and I. K. Schuller, J. Magn. Magn. Mater. \textbf{192}, 203 (1999). 	
\bibitem{Buzdin2005}
	A. I. Buzdin, Rev. Mod. Phys. \textbf{77}, 935 (2005).
\bibitem{Sebas2005}
	F. S. Bergeret, K. B. Efetov, and A. Volkov, Rev. Mod. Phys. \textbf{77}, 1321 (2005).	
\bibitem{Bergeret2012} 
	F. S. Bergeret, A. Verso, A. F. Volkov, Phys. Rev. B {\bf 86}, 060506(R) (2012).
\bibitem{Bergeret2001} 
	F. S. Bergeret, A. F. Volkov, K. B. Efetov, Phys. Rev. Lett. {\bf 86}, 3140  (2001).
\bibitem{Dynes1984}
	R. C. Dynes, J. P. Garno, G. B. Hertel, and T. P. Orlando, Phys. Rev. Lett. \textbf{53}, 2437 (1984).
\bibitem{Pekola2004}
	J. P. Pekola, T. T. Heikkil\"a, A. M. Savin, J. T. Flyktman, F. Giazotto, and F. W. J. Hekking, Phys. Rev. Lett. \textbf{92}, 056804 (2004).
\bibitem{Gamma}
	Unless differently stated, throughout our analysis we set a realistic value for $\Gamma$ of $10^{-4}\Delta_0$ \cite{Pekola2004}. 
\bibitem{Meservey1994}
	R. Meservey and P. M. Tedrow, Phys. Rep. \textbf{238}, 173 (1994).
\bibitem{foot2}
The integrand on the r.h.s of Eq. (\ref{GP}) is even in $\epsilon$  and therefore we restrict the discussion  to  positive energies.
\bibitem{Bergeret2001b} 
	F. S. Bergeret, A. F. Volkov, K. B. Efetov, Phys. Rev. B {\bf 64}, 134506 (2001). 
\bibitem{Ryzanov2001}
	V. V. Ryazanov, V. A. Oboznov, A. Yu. Rusanov, A. V. Veretennikov, A. A. Golubov, and J. Aarts, Phys. Rev. Lett. \textbf{86}, 2427 (2001).
\bibitem{Kontos2001}
	T. Kontos, M. Aprili, J. Lesueur, and X. Grison, Phys. Rev. Lett. \textbf{86}, 304 (2001).
\bibitem{Moodera90} 
X. Hao, J. Moodera, and R. Meservey, Phys. Rev. B \textbf{42}, 8235 (1990).
\bibitem{Moodera08} T. Santos, J. Moodera, K. Raman, E. Negusse, J. Holroyd,
J. Dvorak, M. Liberati, Y. Idzerda, and E. Arenholz, Phys. Rev. Lett.
\textbf{101}, 147201 (2008).
\bibitem{Miao2009}
	G.-X. Miao, M. M\"uller, and J. S. Moodera, Phys. Rev. Lett. \textbf{102}, 076601 (2009).
\bibitem{Giazotto2008}
	F. Giazotto, T. T. Heikkil\"a, G. P. Pepe, P. Helisto, A. Luukanen, and J. P. Pekola, Appl. Phys. Lett. \textbf{92}, 162507 (2008).
\bibitem{Nielsen2002}
	M. A. Nielsen and I. L. Chuang, \emph{Quantum Computation and Quantum Information} (Cambridge University Press, 2002).
\bibitem{Giazotto2004}
	F. Giazotto and J. P. Pekola, J. Appl. Phys. \textbf{97}, 023908 (2005).
\bibitem{Tirelli2008}
	S. Tirelli, A. M. Savin, C. Pascual Garcia, J. P. Pekola, F. Beltram, and F. Giazotto, Phys. Rev. Lett. \textbf{101}, 077004 (2008).
\bibitem{bauer2010} 
	G. E. W. Bauer, A. H. MacDonald, and S. Maekawa, Solid State Commun. \textbf{150}, 459 (2010).
\bibitem{bauer2012} 	
G. E. W. Bauer, E. Saitoh, and B. J. van Wees, Nature Mater. \textbf{11}, 391 (2012).
\end{thebibliography}
\end{document}